\documentclass[conference,compsoc]{IEEEtran}
\usepackage{tikz}
\usepackage{amsmath}
\usepackage{tabularx}

\begin{document}

\title{Towards a Graph Neural Network-Based Approach for Estimating Hidden States in Cyber Attack Simulations}

\author{\IEEEauthorblockN{Pontus Johnson, Mathias Ekstedt}
\IEEEauthorblockA{\textit{KTH Royal Institute of Technology}\\
Stockholm, Sweden \\
pontusj@kth.se}}

% make the title area
\maketitle

\begin{abstract}
This work-in-progress paper introduces a prototype for a novel Graph Neural Network (GNN) based approach to estimate hidden states in cyber attack simulations. Utilizing the Meta Attack Language (MAL) in conjunction with Relational Dynamic Decision Language (RDDL) conformant simulations, our framework aims to map the intricate complexity of cyber attacks with a vast number of possible vectors in the simulations. While the prototype is yet to be completed and validated, we discuss its foundational concepts, the architecture, and the potential implications for the field of computer security.
\end{abstract}

% no keywords

% For peer review papers, you can put extra information on the cover
% page as needed:
% \ifCLASSOPTIONpeerreview
% \begin{center} \bfseries EDICS Category: 3-BBND \end{center}
% \fi
%
% For peerreview papers, this IEEEtran command inserts a page break and
% creates the second title. It will be ignored for other modes.
\IEEEpeerreviewmaketitle

\section{Introduction}
\label{intro}

The increasing sophistication and frequency of cyber attacks necessitate advanced methodologies for understanding and predicting attack vectors in complex network environments. Traditional models often fall short in capturing the dynamic and stochastic nature of real-world attacks. Our research aims to bridge this gap by introducing a novel approach leveraging Graph Neural Networks (GNNs) \cite{zhou2020graph} and dynamic simulation frameworks. The integration of GNN with Meta Attack Language (MAL) \cite{MAL} and in conjunction with Relational Dynamic Decision Language (RDDL) based simulations \cite{sanner2010relational} presents an innovative method to estimate hidden states in cyber attacks, enhancing our understanding and predictive capabilities in cybersecurity.

\section{Background and Related Work}

Recent advancements in machine learning, particularly in GNNs \cite{zhou2020graph}, have shown remarkable results in interpreting complex, graph-structured data, a common representation in network security contexts. GNNs' ability to capture dependencies and interactions between nodes in a network makes them an apt choice for modeling cyber attack scenarios. Prior works in cybersecurity have utilized various techniques, ranging from traditional attack graphs to advanced machine learning models, for threat detection and analysis. However, the combination of GNNs with MAL and RDDL for cyber attack simulations represents a novel approach in the field, offering a more dynamic and realistic representation of potential security breaches.

\section{Methodology}

The methodology underpinning our prototype is founded on integrating Graph Neural Networks (GNNs) with simulations based on MAL and RDDL. This integration aims to address the complexities inherent in cyber attack simulations, particularly in estimating hidden attack states.

MAL provides a formalism for modeling cyber attack scenarios, defining the entities involved, their attributes, and potential attack vectors. It allows for the codification of domain-specific cybersecurity knowledge, facilitating the creation of detailed and nuanced attack scenarios. In our approach, MAL is used to define the structure of the cyber environment, detailing the assets, vulnerabilities, and potential attack paths.

RDDL enables the specification of dynamic systems and decision-making processes, which are critical for simulating cyber attack scenarios with temporal and stochastic elements. Using PyRDDLGym \cite{taitler2022pyrddlgym}, we create RDDL-compliant simulations that mimic the progression of attacks in a network. These simulations consider various factors, including attacker behavior, system defenses, and the probabilistic nature of certain attack outcomes.

The integration of MAL and RDDL provides a comprehensive framework for simulating cyber attacks. MAL defines the static structure of the attack scenarios, while RDDL adds dynamism, allowing for the simulation of how attacks unfold over time. The resulting simulations generate data that reflects the complex interplay of various factors in a cyber attack, forming the basis for training our GNN model.

\section{Observation Logic and Handling of Hidden States}

A critical aspect of our prototype is its approach to managing hidden states in cyber attack simulations. Hidden states refer to the aspects of a cyber attack that are not directly observable, such as the compromised status of a system without explicit evidence. Accurately estimating these hidden states is pivotal for understanding the full scope of an attack and for effective cybersecurity measures. 

In our framework, the observation logic is intricately tied to both the MAL and RDDL components. MAL provides the structural backbone of the attack scenarios, defining the entities, attack steps, and possible state configurations. However, MAL inherently does not deal with the dynamism of state changes or the probabilistic nature of state visibility in a complex network attack scenario. This is where RDDL plays a crucial role.

RDDL introduces a dynamic perspective, enabling the modeling of temporal changes and stochastic behaviors within the system. It allows us to specify how the states evolve over time, under different conditions, and how these changes may or may not be observable. The observation model in RDDL is used to define what information about the system's state is available at any given point in time, which in turn influences the GNN's learning process.

The GNN, trained on the data generated from MAL-defined structures and RDDL-based simulations, learns to infer the hidden states based on the observable data patterns. It utilizes the temporal and stochastic information encoded in the RDDL simulations to make educated guesses about the unobservable aspects of the system. For example, if certain attack steps in the MAL model are known to lead to specific system behaviors, but direct evidence of these steps is lacking, the GNN can estimate the likelihood of these steps being compromised based on the observed data from other parts of the network.

In conclusion, the observation logic in our prototype is a synergistic blend of MAL's structural definitions and RDDL's dynamic and probabilistic modeling, further enhanced by the GNN's advanced learning capabilities. This approach allows for a more comprehensive and realistic representation of cyber attacks, including the estimation of hidden states, which is crucial for effective cybersecurity analysis and response.

\section{Prototype Architecture}

The architecture of our prototype is designed to process, learn from, and predict the states of a simulated cyber attack environment. It comprises the following key components:

Using MAL, we define the parameters of the cyber attack scenarios, including the network architecture, potential vulnerabilities, and attack steps. This definition forms the backbone of the simulation model.

The RDDL-based simulation engine, implemented via PyRDDLGym, dynamically generates attack graphs based on the MAL-defined rules. These attack graphs depict the progression of attacks, including the sequential and conditional relationships between different attack steps.

The GNN model is the core predictive component of our architecture. It is designed to process the output of the dynamic simulations, learning the patterns and relationships within the attack graph. The GNN model aims to estimate hidden states in the attack graph, such as identifying which attack steps might have been compromised without direct evidence.

This module converts the output of the RDDL simulations into a format suitable for GNN training. It ensures that the temporal and relational aspects of the attack graph are accurately represented and fed into the GNN.

\section{Preliminary Implementation}

The preliminary implementation of our prototype includes the initial setup of the MAL model, RDDL specifications, and the basic GNN model.

We have developed MAL models for various attack scenarios, detailing the assets, vulnerabilities, and attack steps relevant to each scenario. Corresponding RDDL specifications have been created to simulate these scenarios dynamically. These specifications account for probabilistic elements and temporal dynamics inherent in cyber attacks.

A basic GNN model has been implemented to process the attack graphs generated by the RDDL simulations. This model is currently capable of ingesting graph data and learning from the structural and temporal patterns present in the simulations. However, further refinement and optimization of the model are ongoing.

The integration of the different components is underway, with a focus on ensuring seamless data flow and processing across the MAL, RDDL, and GNN modules. The development environment is centered around Python.

\section{Challenges and Future Work}
Developing a robust GNN-based framework for cyber attack simulations presents several challenges. Key among them is the integration of the MAL-defined scenarios with the dynamic RDDL simulations, ensuring that the GNN can effectively learn from these complex, multi-faceted inputs. Additionally, refining the GNN to accurately predict hidden states in a network, especially under varying and unforeseen attack conditions, requires extensive research and iterative testing. Future work will focus on completing the system integration, improving the GNN model's accuracy, and conducting comprehensive validation exercises. We also aim to extend the framework's capabilities to include a wider range of attack scenarios and network configurations, enhancing its applicability in real-world settings.

\section{Conclusion}
This paper presents an early-stage prototype that employs a novel GNN-based approach for estimating hidden states in cyber attack simulations. By integrating MAL-defined scenarios with RDDL-based dynamic simulations, our framework offers a promising new direction in cybersecurity research. While still in its preliminary stages, the potential of this approach to provide more dynamic, accurate, and realistic simulations of cyber attacks is significant. As we continue to develop and refine this prototype, we anticipate that it will contribute substantially to the field of cybersecurity, offering enhanced tools for understanding and mitigating cyber threats.

\section*{Availability}
%-------------------------------------------------------------------------------

A prototype implementation is available on GitHub at https://github.com/pontusj101/gnn\_ids\_trainer.

%-------------------------------------------------------------------------------
\bibliographystyle{plain}
\bibliography{main}

% that's all folks
\end{document}